# Review and Test of Steganography Techniques

Joab Kose, Oscar Bautista Chia, Vashish Baboolal
EEL 6803 – Advanced Digital Forensics
Florida International University
10555 W Flagler St Miami, FL 33174

*Abstract*— Steganography is the art of concealing a secret message within an appropriate-multimedia carrier such as images, audio, video files, and even network packets. Steganographic techniques have been used since ancient times to hide the message from third-parties deemed to be enemies. On one hand, Steganography is a useful technique that has been applied in various useful applications. On the other hand, however, the same technique has been applied and used for the wrong purposes by people who have ill intentions. The increase in computational power and increase in security-awareness over the past few years have propelled the application of steganography in computational security-techniques. Being robust, undetectable, and having a good capacity of hidden-data, steganography has been preferred for hiding data than watermarking and cryptographic-techniques. This paper clarifies the application of steganography on different multimedia-carriers by using various potential tools, methods, and principles to either apply or detect steganography techniques.

## I. INTRODUCTION

What is Steganography? A steganographic system consists of two main components: Encoder and Decoder. The Encoder is responsible for embedding the secret message or confidential information within the cover medium or covert channel. The Decoder, on the other hand, is responsible for decrypting or retrieving the secret message or secret file from an encrypted file, image, or physical object through the use of a secretly shared key between the sender and intended receiver.

Another component of a steganographic system is the medium that is used to hide or embed the sensitive information. These mediums also referred to as "cover" files can be images, documents, music, or any type of digital media files as well as network packets as well. The result of the "cover" file after the steganography process or after the information has been embedded within is referred to as "stego-file".

## II. ORIGIN OF STEGANOGRAPHY

How did Steganography originate and evolve into usage in the present-day cyber world? Steganography is not a modern technique of data hiding as it has its roots in the ancient past. Historically, steganography took on physical forms long before the dawn of the computer age to send secret data. Messages were hidden on tangible objects and in some countries within humans as well.

"In 5th century BC, Histaiacus shaved the head of a messenger, then wrote the secret message on his bald head and waited for the hair to grow back to send the messenger to the other party. To retrieve the message, his hairs were shaved again. This method was very time-consuming." [1].

"Gaspar Schott (1608-1666) wrote the book named as "Schola Steganographica," in which the technique to hide messages in music scores has been discussed. Here, each letter of the message corresponds to one note. The 'Ave Maria' code, originally proposed by Johannes Trithemius (1462-1516), is also expanded in this book in forty tables and each table consists of twenty-four entries in four languages. Each letter was replaced by the word in the corresponding table entry and thus the stego text was prepared. It has been discovered that these forty tables can be translated by reducing them…and applying reverse alphabets." [2].

Famous Composer J.S. Bach hid data in music scores based on the number of occurrences of notes. Messages have been hidden in geometric drawings using points and line ends.

"The use of Ciphers for hatching "The Babington Plot" in March 1586 to assassinate Queen Elizabeth and put Mary, Queen of Scots, a Catholic, on the English throne led to the imprisonment and subsequent execution of Mary." [3].

"A security protocol was developed by ancient China in which the sender and the receiver had the same paper mask having several holes at random locations. The sender could write the secret message into the holes by placing his mask over a paper, remove the mask and compose a cover message. The receiver could get the secret message by placing his mask over the letter received." [4].

In the 17th century and later on, invisible ink was used to print very small dots instead of making holes. Invisible ink, made up of organic substances i.e. milk or salt ammoniac dissolved in water and developed heat which helped in steganography, but this technology did not succeed because of the invention of "universal developers". These could determine the parts of paper being wet from the effects on the surface of fibers.

Several steganographic techniques had been used during World War II by the German Nazis. Steganography in the modern era has been elevated to incredible levels of sophistication. The advent of Computer Systems and their accompanying hardware and software have become the backbone of all the in-depth work. Computer systems employ media such as images, audio, video, text, all of which can be used and manipulated digitally to perform steganography.

Earlier, steganography was implemented using some physical medium within tangible objects. Today, it is implemented electronically by using several intangible objects. Data can be hidden using any type of media, be it images or music files, video clips, text files, SMS, etc. Unlike watermarking and cryptography, the main focus in steganography revolves around concealing the existence of any secret communication taking place.

The main goal of steganography is to secretly communicate a message between a sender and its intended recipient and to also conceal that a message is being secretly transmitted as well.

In steganography, we are not concerned about the security of sensitive information and whether it is altered or not but only about how the message is hidden for covert transmission.

## III. TECHNIQUES FOR STEGANOGRAPHY

*A. Image Steganography*

Image Steganography is the most popular type of steganography. The scope of image steganography is large because of the various image formats available such as BMP, JPEG, PNG, GIF, etc. The user can opt from one of these image formats as required. Different steganographic techniques have been developed based on these different image formats.

In general, a digital image is an arrangement of small dots known as pixels (picture elements), each having different light intensity. "The bit depth is the number of bits in a pixel. The smallest bit depth for color images is 8, which means 8 bits are used to describe the color of each pixel. Thus, 8-bit depth color and grayscale images can display 256 (i.e. 2-bit depth) different colors or shades of grey respectively. A 24-bit color image can display up to 16,777,216 (224) discrete combinations of Red, Green, and Blue values. These images use the RGB color model which is also known as the true color model. Here, every 8-bits of 24 bits represent one of the three color components i.e. red, green and blue." [5].

Image steganography is divided into two general categories, each of them consisting of different embedding techniques: Spatial Domain Steganography and Transform Domain Steganography [16].

*1) Spatial Domain Steganography*
*a) LSB Method*

This technique involves changing the least significant bit of each color represented by a bit of the secret message, a 24-bit RGB image uses 8 bits for each color, therefore 3bits /pixel can be used to embed a secret message. This technique, though very elementary does have a drawback in the sense that the "cover" file or image must be large enough to embed the secret data otherwise there is the risk of retrieving incorrect information at the decoding stage.

Special attention needs to be exercised when applying this technique to GIF images because the color is specified as an index of a palette of colors which can result in a noticeable color change. To apply this technique in GIF images, it is better to use 8-bit grayscale pictures, which specifies 256 different shades of gray that a small change is hard to detect.

*b) Masking and Filtering*

In this technique, the sensitive information is embedded by altering the pixels in such a way that no noticeable differences can be seen by the human eye. "This method is more robust than LSB in many ways like compression, cropping, and various image processing as it only uses the visual aspects of the cover image" [1].

This method works better in 24-bit and grayscale images.

*2) Transform Domain Steganography*

The message is hidden behind the cover image by modulating coefficients in the frequency domain such as Discrete Cosine Transform (DCT), Discrete Fourier Transform (DFT), or Discrete Wavelet Transform (DWT). The most sophisticated of steganography image embedding, the various data embedding techniques based on Transformations include algorithms resistant offering a very good capacity for steganographic messages – "12.8% of the steganogram's size. After quantization, this algorithm skips all coefficients of value 0 or 1 and replaces the LSB's of the rest of the frequency coefficients by the secret message." [6].

*B. Digital Image Formats*

*1) Audio*

The use of steganography in audio media is less popular than using images because audio files are larger. Some methods for embedding content in an audio file are [17]:

*I. LSB Coding*

As in images, the least significant bit of the cover file is replaced but bits from the embedded media.

*II. Parity Coding*

This technique takes into consideration dissimilarities in parity bits in the cover file to change a bit for those of the embedded media.

*III. Echo Data Hiding*

In this technique, echo sounds are added to the cover media, data is hidden in terms of initial amplitude, decay rate, and delay [17]. The amplitude determines the original sound amplitude, the decay rate is used for the determination of the echo function and the delay specifies the interval between the original sound and the echo sounds.

*2) Video*

Videos offer more channels to hide information as video files include text and audio too, each of those can be used to be applied specific techniques for that media.

The process of hiding information in the visual portion (the video itself) involves converting it into a sequence of image frames. Next, masking and filtering techniques are applied to analyze the image and determine which portions are appropriate to hide information. This makes difficult the steganalysis process as a third party doesn't know in which frame(s) content is hidden [17].

*3) GIF*

GIF, short for Graphics Interchange Format is a bitmap image format. The GIF format supports up to 8-bits per pixel, thus allowing a single image to reference a palette of up to 256 distinct colors, chosen from the 24-bit RGB color space. Popular usage for websites.

*4) BMP*

BMP: This image format is also known as a bitmap image file format common for MS Windows. The BMP images are large and their quality varies from medium to high.

*5) JPEG*

JPEG: The term "JPEG" is an abbreviation for the Joint Photographic Experts Group. JPEG is a commonly used file format of lossy compression for the digital image. This is the most widely used image format for photographic images. JPEG images are of high quality and small in size.

## IV. STATISTICAL & SIGNATURE STEGANALYSIS

Steganalysis is the technique to detect the presence of some secret message behind the suspected cover medium. It can be said that steganalysis breaks the purpose of steganography by detecting the existence of some hidden message and thereafter destroying it. Steganalysis can be classified into two categories: Statistical and Signature Steganalysis.

*A. Statistical Steganalysis*

Various statistics of the cover file also change when some data is hidden behind it. Statistical steganalysis involves mathematical computations to detect the presence of hidden information. As mathematical observations are more accurate than visual observations, this method is more powerful than the signature steganalysis

*B. Signature Steganalysis*

The properties of the cover file also change when some secret message is embedded behind it. Due to such changes, unusual patterns can be detected in the cover file and these patterns are known as signatures. Signature steganalysis looks for such patterns to conclude the existence of the hidden message. Though steganography techniques hide content behind the cover file in a way to remain unnoticeable to the human eye, due to such signatures, users can suspect some cover file.

## V. STEGANOGRAPHY TECHNIQUES

*A. Images*

Since image files can be of various formats, such as BMP, JPEG, or GIF, different tools are necessary to perform steganography. For JPEG images, JPHide or JPseek are used to alter the image slightly [12]. How these programs work is that they take a lossless cover-image and the message to be hidden, and generate a stego-image. These tools are used with steganography performed in the transform domain, such as discrete cosine transform (DCT) [13].

DCT has two properties that allow it to facilitate steganography—decorrelation and energy compaction [14]. Image transformation removes redundancy between pixels, which means that uncorrelated transform coefficients can be encoded independently [14]. Energy compaction is "the ability to pack input data into as few coefficients as possible," which allows for coefficients to be discarded without severe visual distortion [14]. These two properties together allow for steganography in JPEG images by rounding DCT coefficients up or down for individual bits to be embedded [13].

Instead of altering an image, another steganography technique that is used with image files is to manipulate the image's color palette, leaving the original image intact [12]. This technique is most often used on GIF images with the tool Gifshuffle [12].

Another technique for hiding image files within another file is through the use of the "copy" command in Windows Command-Line. For example, the following command takes a group of images or files that are zipped into a folder and hides it within another cover image: **"copy /b imagestohide.zip+coverimage.jpg stegoimage.jpg".** The "copy" command with the "/b" flag instructs Windows to read the files in as a byte stream and copy them byte by byte to the stegoimage.jpg. Opening both cover images and stego images, no noticeable visual differences would be seen but using an unzipping tool such as Winrar on the stego image would extract the zipped folder and its contents.

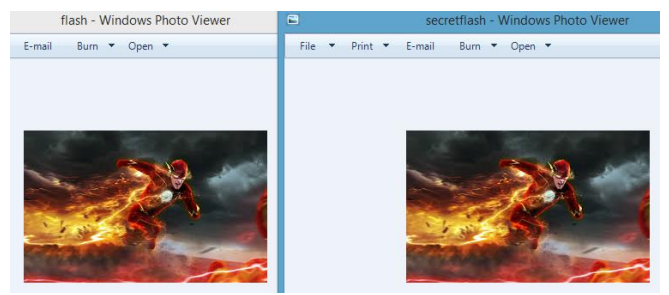

Figure A - Hiding text within Image

The previous technique can also be used to embed a secret text message within an image file. It will use the same command and the name of the file that contains the secret message for example "secret.txt", which would replace the "imagetohide.zip" part of the previous command as shown in Figure A above. Again no distinguishable differences would be seen between both the original cover image and the final stego image as shown in Figure B. To retrieve the hidden message in this approach, the stego image can be opened with the Notepad application, and the secret message would be found towards the end of the file as can be seen in Figure C.

Figure B - Comparing Cover and Stego Images.

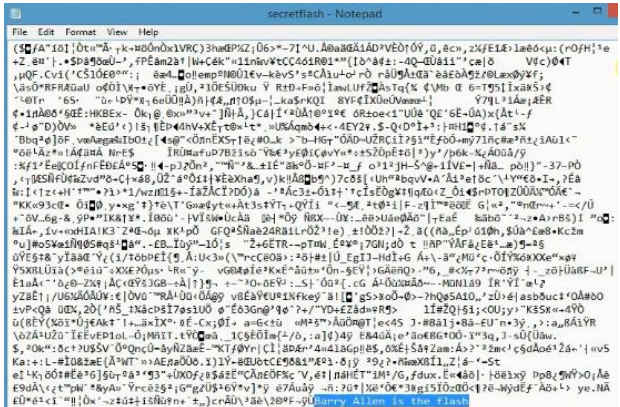

Figure C - Recovered Secret Message

The command-line approach of hiding files or text within other files may be a little tedious for some users but it is far more discreet than installing actual applications such as JPHide and JPSeek as the installation of applications would be recognized within the system's registry whereas the Windows Command Line comes built into the Operating System already.

*B. Audio*

The use of steganography in audio media is less popular than using images because audio files are larger [A]. Three tools are used with audio files if the stenographer chooses to hide information in audio. MP3Stego is used to hide information in MP3 files, commonly as a watermark [12]. StegoWav takes advantage of the Windows Wav format to hide information. Steghide is used with the least significant bit method in the audio carrier medium [12].

However, more discreet everyday graphical music editing tools such as Coagula can also be used to perform steganography attacks, and Audacity to retrieve the hidden messages.

Coagula is a graphical editor that constructs sounds by analyzing the sine-waves within images. Each line within an image controls the amplitude of one oscillator. For instance, the vertical position of a single-pixel determines the frequency while the horizontal position determines the length of time. Furthermore, the colors used in the pixel themselves contribute to different sound effects. Red and green colors relate to the stereo placement of the sound such as left channel and right channel components while the bright color results in a brighter sound and vice-versa. Using Bitmap images created with a photo editing tool such as Photoshop or simply Paint, Coagula creates an audio file in .wav format by analyzing the pixel values within the image. The tool also provides additional features to modify or add different textures to the image thus creating various sound effects.

In Figure D below, an image was created using the Paint application and contained the secret message "The password is 1234567890", written in white text over a black background. The reason for these color options in text and background is so that that message can be seen easily when the wave file is observed under the Spectrogram view. Once the image-message was created and saved in Bitmap format, it was then opened in Coagula to add additional sounds effects noticeable by the reddish-green scribbles on the image. Any pattern or effect could have been added to the image to create various audio effects and improve the disguise of the stego wave file.

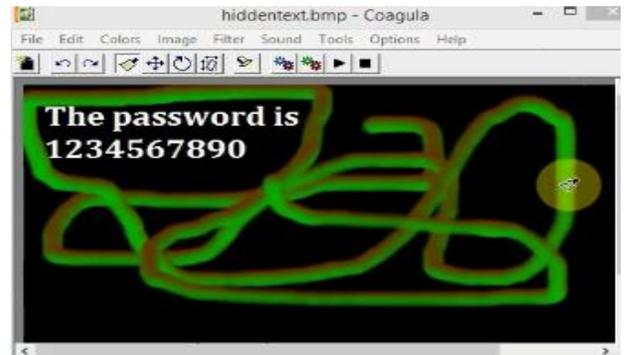

Figure D – Adding Sound Effects to Image Message

Once the audio WAV file has been created with Coagula, to a regular user it can listen as a regular music file. However, its intended recipient would be able to retrieve the embedded message using the Audacity tool. Audacity is a simple digital audio editor for Windows and used for audio editing and recording tasks and is therefore very discreet as well, unlike actual steganography tools. Figure E below shows the recovered message after using the Spectrogram feature within Audacity.

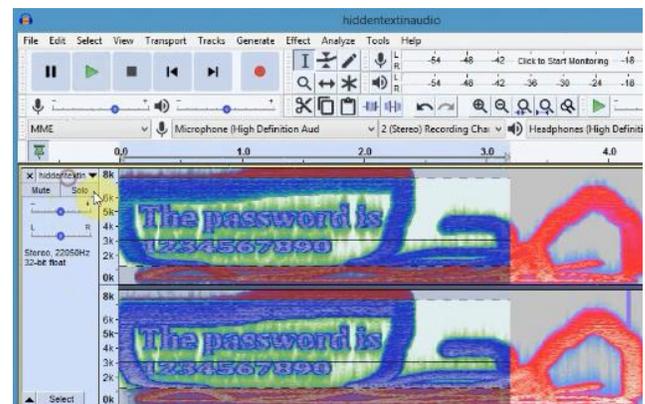

Figure D – Spectrogram in Audacity

*C. Text*

Steganography can be used in text files with the use of StegParty, which uses a set of rules based on the flexible rules of the English language to hide small amounts of data that appears as a small typo or a small grammatical error [12]. Although the amount of data that can be hidden is small, it can still be very effective. Because of human error, people typically ignore or overlook typos or grammatical errors, which makes the presence of these items inconspicuous. This tool is not limited to only text files such as Word documents, it can also be extended to be used with HTML files.

In the specific case of formatted text, such as an MS Word document, the formatting characters increase the capacity to conceal information, for example, by changing the font color of space characters which does not make any change to the document visualization.

Furthermore, tools such as Unisteg by Irongeek can be used to hide text within other text. The tool works by encoding the Unicode characters of the secret text in such a way that it appears normal to the human eye but in actuality, they have a different value as shown in Figure E next. The secret message is encoded onto the cover text to produce a stego text with almost normal Unicode text and no visible differences between itself and the cover text.

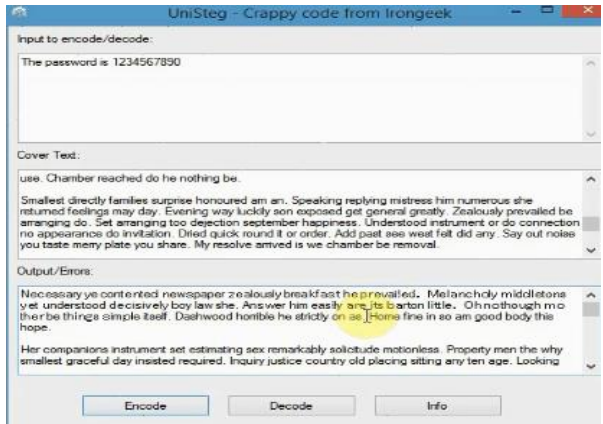

Figure E – Hiding text within the text with Unisteg

*D. Network Packets*

Another method of communicating messages secretly across is the network is within the network itself through the use of the TCP/IP protocols which are what our network communication model is built upon. By exploiting some of these protocols covert channels can be created so that information can be secretly transmitted between hosts. By using this technique, message injected packets or stego packets can bypass packet filters and network sniffers and not be detected thus ensuring communication and transmission of messages.

TCP packet headers offer the most possibilities for covert communication due to the many fields within which information can be hidden. For example, there is a Sequence Number field which signifies what TCP stream a particular packet belongs to [19]. Having a size of 32 bits, it is ideal for hiding information within this field as the Sequence Numbers can be converted to ASCII characters by dividing by 16777216 [20].

Another field within the TCP header that can be exploited is the Acknowledgement Number Field or the ACK field. It involves spoofing the sender's IP address to the recipient's address so that when the sender transmits an encoded message, it will get sent directly to the receiver without undergoing an SYN/ACK process by the server.

The IP header packet can also be used to create a covert channel for communication. It contains 14 fields of which 13 are required and the 14th being optional and exploitable for secret messages. [19]. However, a more discreet approach would be to target the Identification field instead of the optional field as the ID field is mandatory. The ID field is 2 bytes long and provides a unique number for packet reassembly purposes. Therefore, a steganography attack can be done by exploiting this field by replacing the ID number with the numerical ASCII representation of the character of the message to be transmitted.

A tool that can be used to carry out this feat is Covert_TCP which is a C program that allows for messages to be secretly transmitted over a network by encoding the message within either the IP ID header, Sequence number field, or the ACK field.

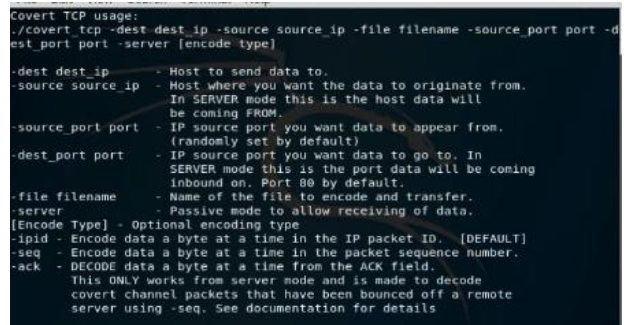

Figure F – Covert_TCP

From Figure F above we can see that certain parameters need to be set when running the command. Once the destination and source addresses have been set along with the type of encoding and the file containing the secret message, the program will begin the communication channel and transmit the secret message. As in the case of IP ID encoding, the secret message will be transmitted one byte at a time to the receiver.

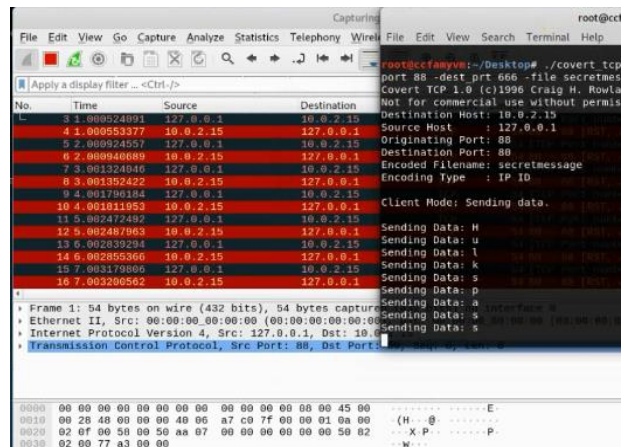

Figure G- Observing Covert_TCP in Action

As seen in Figure G, Covert_TCP transmits the secret message to the receiver one byte at a time as it places the message a byte at a time into the Identification field. Using the Wireshark tool we were able to observe this communication channel and identify the secret message being encoded and transmitted within the IP header packets. Therefore, the intended receiver of the message can simply use Wireshark to observe the network traffic from the sender and retrieve the secret message successfully.

*E. Graph Steganography*

As opposed to the previous techniques to apply steganography to digital media, where a secret message is embedded in a cover media, graph steganography transforms the secret message into data that can be used to plot a graph with a made-up purpose, so that it passes inadvertently to check-ups for suspicious data.

One of such implementations by F. Akhter [18] uses Huffman coding to assign a numeric value to every word in a secret message, then applies a scaling factor to add an additional layer of security.

A brief description of this technique is provided hereunder, a complete explanation of the application of this technique can be found in [18]:

Huffman coding assigns a binary code to each letter in a text-based on its frequency of appearance, giving them more frequent letters a shorter binary representation while letters with lower frequency will be assigned the longest binary codes. For example, the string "this is a test" has this letter frequency:

| Letter    | a | e | h | i | s | t |
|-----------|---|---|---|---|---|---|
| Frequency | 1 | 1 | 1 | 2 | 3 | 3 |

After building a Huffman tree and obtaining from it the binary representation for each letter, one possible result is:

```
t:  11
s:  10
i:  00
h:  010
a:  0110
e:  0111
```

The next step is to obtain the binary representation of a word from the letter's binary code, this is done by simple concatenation, for instance, the word "is" would be 00 10 where the first two binary digits correspond to the letter "i", and 10 corresponds to the letter "s". After that, an additional step not mentioned in the referred paper is to add a leading "1" to obtain 10010, so that the decimal value would be 18, instead of 2 which is the original case. The purpose of this leading one is to be able to count for the leading zeros that would be lost when the receiver converts the decimal "2" to binary "10"; by adding a leading "1", the receiver will still get 10010 as the binary for 18 and then it is just needed to remove the leading "1" before using the same Huffman table to obtain the letters for each word from the word's binary representation.

Lastly, to add a layer of security, the author adds an α value which corresponds to the space character as well as a multiplier β value. Both α and β are known to transmitter and receiver to be able to retrieve the message correctly.

## VI.  APPLICATIONS OF STEGANOGRAPHY IN DIGITAL MEDIA

### A. Watermarking

It must be noted that there is an important difference between watermarking and steganography, although steganographic methods and techniques are used in watermarking. The goal of steganography is to conceal information and watermarking only extends the data of the file [15]. The application of steganography in watermarking is that certain pixels in the image are masked so that they are a certain percentage lighter or darker than the rest of the image, but the data of the image is not altered. Another difference between watermarking and steganography is that watermarking is used to intentionally identify a piece of media as copyrighted material. Information that is hidden through steganography is typically intended to be concealed in such a way that it does not appear that there is any hidden message [12]. The way that watermarking is still considered an application of steganography is that although there is a visible change to the media, the embedded information does not undergo a significant change [12].

### B. Privacy

Although not a substitute for encryption, steganography can be used to send information without being monitored. The difference between true encryption and steganography is that encryption encodes data so that unintended recipients cannot comprehend the original meaning, while steganography is used to prevent unintended recipients from realizing that there is any additional data being sent [12].

Steganography can become an important tool in situations where cryptography and strong encryption are outlawed [15]. The morality of whether or not the use of steganography as a tool for encryption is good or bad is dependent on the intent of use. Steganography can be used to encrypt private conversations, or even be potentially used to protect national security [12]. There is always the risk that steganography can be used to traffic illegal material through digital media on the web, so it is important for law enforcement to understand how to detect steganography [15].

### C. Corporations

Other than using watermarks to protect intellectual property, corporations also need to protect trade secrets. Trade secrets include company-specific methods or recipes that provide a competitive advantage over other companies of the same industry. To avoid this kind of information leak, it is important for company employees to avoid accidentally leaking information by communicating through common media [12]. This problem can be solved in situations where trade secrets must be communicated if steganography is used to hide portions of the trade secret in text-based or image-based media.

## VII.  LIMITATIONS OF STEGANOGRAPHY

Similar to how security techniques such as encryption is limited, so to as steganography. Primarily, both the sender and intended recipient of the message must first meet or communicate secretly in order to agree upon which steganography technique to use in order to communicate their messages secretly in public.

Secondly, the amount of information that can be secretly hidden is limited by the size of the cover media itself otherwise the message would be affected. Ultimately, "the less limitations that exist on the integrity of the medium, the more potential it has for hiding data" [21].

## VIII. Conclusion

Digital steganography exhibits endless possibilities, the numerous techniques can only be classified in general groups such as media type (text, audio, image, video, protocol, etc.) and by type of technique from the simplest modifications, or use of spare fields to complex transformation and processing that discourage an attacker to try to find any message, and focusing the efforts filtering or transforming the stego-media structure in a way that whatever message is hidden cannot be recovered, in other words thwarting the communication.

The more complex the technique used to embed secret information into a cover media, the more variations could be derived. For example, the LSB technique is simple and variations can come by changing the order that the embedded media should be retrieved, and therefore a previous agreement should exist between sender and receiver, in contrast, transform space techniques are not only more complex but varied as well.

The main characteristic that makes steganography so powerful and therefore dangerous, is it hides the existence of a message in the first place, for good or for bad, a covert communication may exist for a long period before any small suspicion is raised and communications start being analyzed, also different and improved steganography method that is not yet known to the general public may already exist, and are being exploited in scenarios such as to resist jpeg image compression.

Steganography can be combined with cryptography techniques, and indeed this is what is done when the parties deal with sensitive information, this approach makes it practically impossible for a third party to discover any piece of information without knowledge about the encryption and steganographic system: in the case that an embedded message is discovered and they can assure that the encrypted message is indeed and encrypted content and not garbage information, they would still need a decryption key to retrieve the hidden content.

Steganography can be used positively, via digital watermarking and fingerprinting techniques, to protect digital intellectual property [9].

Steganography and steganalysis are a powerful tool that can be used in cyber-combat scenarios between nations.